# Solar energy conversion and light emission in an atomic monolayer p-n diode

Andreas Pospischil, Marco M. Furchi, and Thomas Mueller[*]

*Vienna University of Technology, Institute of Photonics,*

*Gußhausstraße 27-29, 1040 Vienna, Austria*

**Two-dimensional (2D) atomic crystals[1], such as graphene[2] and atomically thin transition metal dichalcogenides[3, 4] (TMDCs), are currently receiving a lot of attention. They are crystalline, and thus of high material quality, even so, they can be produced in large areas and are bendable, thus providing opportunities for novel applications. Here, we report a truly 2D p-n junction diode, based on an electrostatically doped[5] tungsten diselenide ($WSe_2$) monolayer. As p-n diodes are the basic building block in a wide variety of optoelectronic devices, our demonstration constitutes an important advance towards 2D optoelectronics. We present applications as (i) photovoltaic solar cell, (ii) photodiode, and (iii) light emitting diode. Light power conversion and electroluminescence efficiencies are ≈ 0.5 % and ≈ 0.1 %, respectively. Given the recent advances in large-scale production of 2D crystals[6, 7], we expect them to profoundly impact future developments in solar, lighting, and display technologies.**

---

[*] Email: thomas.mueller@tuwien.ac.at

Most of today's electronic devices rely on bulk semiconductor crystals. However, their rigidity, heavy weight, and high costs of production hinder seamless integration into everyday objects. Therefore other, non-crystalline, materials are currently investigated, with organic and thin-film semiconductors[8, 9] being the most prominent ones. These are, however, notoriously known for their low material quality and degradation over time. 2D atomic crystals, on the other hand, are crystalline, yet they can (potentially) be produced at low cost and in large areas, making them attractive for applications such as solar cells or display panels.

P-n junction diodes are an integral part of many optoelectronic devices. P–n junctions have previously also been formed in graphene[10], but did not show diode-like rectification behavior, due to Klein-tunneling[11]. Although graphene can be employed for photodetection[12], it does not produce a sizable photovoltage because of its zero bandgap. For the same reason, graphene p-n junctions would not produce any electrically driven light emission. Graphene has extensively been explored for solar and display applications, but only in conjunction with other materials[6, 13–15]. More recently, other 2D crystals, such as $MoS_2$ and $WSe_2$, have gained increasing attention[3, 4, 16–19], as these materials have a bandgap. Bulk $MoS_2$ and $WSe_2$ are indirect semiconductors, whereas their monolayers exhibit a direct gap[20, 21], making them attractive for optoelectronics. Very recently, a p-n junction diode has been realized[22] in ionic liquid gated bulk $MoS_2$. However, to our knowledge, a monolayer p-n diode has not yet been demonstrated in any 2D crystal.

In our device, electrostatic doping is used to form a monolayer $WSe_2$ lateral p-n junction diode. As schematically illustrated in Figure 1a, split gate electrodes couple to two different regions of a $WSe_2$ flake (crystal structure in Figure 1b). Biasing one gate with a negative and the other with a positive voltage, draws holes and electrons, respectively, into the channel and a p-n junction is realized. The device can also be operated as a resistor by applying gate voltages of same polarity. A similar concept has previously been employed to realize carbon nanotube diodes[5] and double-gated organic field effect transistors[23]. Details of the device fabrication are outlined in the Methods section. A microscope image is shown in Figure 1c.

Prior to device fabrication, the $WSe_2$ flake was extensively characterized to assure monolayer thickness. This is essential for achieving efficient electrically driven light emission, as only monolayers exhibit a direct bandgap[20, 21]. In Figure 1d we show photoluminescence (PL) from mono-, bi-, and multi-layer flakes. In agreement with previous reports[24], pronounced PL emission at 1.64 eV, with spectral width of 56 meV (full-width at half-maximum – FWHM), is obtained at the direct excitonic transition of monolayer $WSe_2$. As the thickness is increased, a strong reduction in quantum yield is observed due to the transition of the material from being a direct to an indirect semiconductor[20, 21, 24]. In addition to PL, we performed Raman measurements, the results of which are presented for the monolayer flake in the inset of Figure 1d. We find a pronounced peak at 248.7 $cm^{-1}$ and a small shoulder at 260 $cm^{-1}$, proofing that the flake is indeed monolayer[24].

After device fabrication, we acquired in a first step the gating characteristics by interconnecting the two gate electrodes ($V_{G1} = V_{G2}$) and leaving the substrate electrically floating. As depicted in Figure 2a, the device exhibits clear ambipolar transfer characteristics, demonstrating that both electrons (n) and holes (p) can be injected into the channel. This is in contrast to MoS$_2$, where strong Fermi level pinning[25] and the large bandgap make it more difficult to obtain p-conduction, which has as yet only been achieved by means of ionic liquid gating[26]. The rather low on/off ratio of 1–5×10$^2$ is attributed to inefficient gating of the intrinsic channel region, which occurs only by stray fields from the split gates. For the same reason it is not possible to extract any meaningful field effect mobilities from this measurement. Mobilities, obtained for standard back-gated field effect transistor are in the range 0.1–1 cm$^2$/Vs, and we expect similar values in the diode device. The transfer characteristic shows a small hysteresis that we attribute to charge traps in the gate dielectric[27] and/or interface water[28].

Figure 2b shows an idealized band diagram of the device when operated as p-n junction diode ($V_{G1} < 0$, $V_{G2} > 0$). From this figure, our motivation for using asymmetric contact metallization becomes apparent. At the anode we used palladium (Pd), which is a high work function metal, to align the Fermi level with the valence band edge of WSe$_2$ for hole injection[18, 29]. At the cathode we used titanium (Ti), which has low work function, for electron injection. However, as Ti is prone to oxidation, other metals, such as nickel[29], may be preferable.

In Figure 2c we present electrical characteristics measured under different gate bias configurations. The dashed lines show I-V curves when operating the device as n- ($V_{G1} = V_{G2} = 40$ V; green dashed line) or p-type ($V_{G1} = V_{G2} = -40$ V; blue dashed line) resistor. The n-type resistance is more than order of magnitude smaller than the p-type resistance, in agreement with the gate characteristic. By applying gate biases of opposite polarity ($V_{G1} = -40$ V, $V_{G2} = 40$ V; solid green line), a p-n junction diode is realized and the device now clearly shows rectifying behavior. We deduce the diode parameters – saturation current $I_S$ and ideality factor $n$ – from the Shockley diode equation $I = I_S(\exp(V/nV_T) - 1)$ in the presence of a series resistance $R_S$ that is associated with the contacts and doping regions, and obtain $I_S$ = 0.02 fA, $n$ = 2.6, and $R_S$ = 95 MΩ (see Supplementary Information (SI); $V_T$ denotes the thermal voltage). The large ideality factor implies that recombination current dominates over diffusion current, indicating a large density of trap states in WSe$_2$ that act as recombination centers. We can now reverse the gate voltages ($V_{G1}$ = 40 V, $V_{G2}$ = -40 V; solid blue line) and operate the diode in the opposite (n-p) direction. The higher on-resistance ($R_S$ = 8 GΩ) is attributed to the asymmetric device structure that favors current flow in the other direction.

Figure 3a shows I-V curves under optical illumination (see Methods for experimental details). The meaning of the curves is the same as in Figure 2c. When biased in p-n (n-p) diode configuration, the I-V characteristics are being shifted down (up) and there is a current flow to an external load. Our atomic monolayer diode can thus be applied for photovoltaic solar energy conversion. Importantly, the I-V curves are barely affected when the device is gated as n- or p-type resistor. This is a clear indication that the photoresponse does not arise from one of the Schottky contacts, as it relies on the existence of a p-n junction. Moreover, the photocurrent changes sign when the gate polarities are flipped, which cannot be explained by the built-in potential due to asymmetric contact metallization[30], either.

As the device produces both a current and a voltage, electrical power, $P_{el}$, can be extracted. In the inset we plot $P_{el}$ versus voltage under p-n configuration and incident illumination of 1400 W/m². A maximum electrical output power of $P_{el,m}$ = 9 pW is obtained at $V_m$ = 0.64 V. The corresponding current is $I_m$ = 14 pA and the dashed rectangle in Figure 3a shows the associated power area. For the fill factor, defined as the ratio of maximum obtainable power to the product of the open-circuit voltage, $V_{OC}$, and short-circuit current, $I_{SC}$, a value of $FF = P_{el,m}/(V_{OC}I_{SC}) \approx 0.5$ is obtained. We can now also give an estimate of the power conversion efficiency, which is the percentage of the incident light energy that is converted into electrical energy, $\eta_{PV} = P_{el,m}/P_{opt}$. If we assume that the power conversion takes place in the 0.46 × 2.8 μm² large intrinsic device region, we obtain $\eta_{PV} \approx 0.5$ %. This value is comparable to efficiencies reported[31] for conventional bulk $WSe_2$ p-n junctions ($\eta_{PV} = 0.1$–$0.6$ %). To our knowledge, this constitutes the first demonstration of efficient photovoltaic energy conversion in a 2D atomic crystal. The ≈ 95 % transparency of the $WSe_2$ monolayer (see SI) makes it attractive for semi-transparent solar cells. Moreover, by choosing a proper $WSe_2$ thickness, the tradeoff between optical transparency and efficiency may be adjusted according to the application requirements.

Besides the vertical shift of the I-V under illumination, we observe a slope of the curve at short circuit ($V = 0$). Following the common practice in literature, we model this slope by a shunt resistance $R_{SH} = dV/dI\ |_{V=0} = 100$ GΩ. $R_{SH}$ is mainly associated with carrier recombination loss and reduces the $FF$ by $R_{CH}/R_{SH} \times 100$ % = 37 %, where $R_{CH} \approx V_{OC}/I_{SC} = 37$ GΩ is the characteristic solar cell resistance (see SI). This suggests that there is room for further improvement by improving the material quality. Resistive losses due to $R_S$ can be modeled in similar fashion and are found to be negligible. $I_{SC}$, displayed in Figure 3b, follows a power-law $I_{SC} \sim P_{opt}^{\alpha}$ with $\alpha$ close to one, indicating that the carrier loss is dominated by monomolecular recombination, most probably via disorder-induced trap states[32]. Since $I_{SC}$ is proportional to $P_{opt}$, $V_{OC}$ scales with $\ln(P_{opt})$, as shown in Figure 3c. *FF* is plotted on the right axis of the same figure and is approximately independent of light intensity. The drop in efficiency (also shown in Figure 3c) for weak illumination is due to the power dependence of $V_{OC}$.

When biased in reverse direction, our device can also be operated as a photodiode. A photocurrent of 29 pA is obtained at -1 V, which translates into a photoresponsivity of $R$ = 16 mA/W, or, by taking the ≈ 5 % absorption of a $WSe_2$ monolayer into account, ≈ 0.32 A/W internal responsivity. When operated as resistor, our device also shows a photoconductive response[33] for both n- and p-type conduction (not presented). However, this photoconductor suffers from high dark current and Johnson noise, large power consumption, and slow frequency response, making the diode operation mode more promising for applications.

Next, we present electrically driven light emission from the p-n junction. Figure 4a shows emission spectra recorded by applying gate voltages as shown in the inset, and running constant currents of $I$ = 50, 100, and 200 nA, respectively, through the device. Under reversed (n-p) diode operation we were not able to drive such large currents. The estimated electroluminescence (EL) efficiency, being defined as the ratio of emitted optical power to electrical input power, is $\eta_{EL} \approx 0.1$ % (see SI). Currently, $\eta_{EL}$ is limited by resistive losses in $R_S$ and by non-radiative recombination in the $WSe_2$. It can hence be increased by reducing contact resistance[18, 19] or by using a crystalline substrate[34] to reduce the density of disorder-induced recombination centers[32]. EL has recently also been obtained[35] in monolayer $MoS_2$ by a unipolar current that generates excitons via impact excitation[36] ($\eta_{EL} \approx 0.001$ %). In contrast, our device is operated as a true light emitting diode with ambipolar carrier injection, and we can exclude hot carrier effects for the following reasons: The EL (i) is absent under unipolar operation (green curve in Figure 4a); (ii) exhibits linear current dependence (inset in Figure 4a), whereas impact excitation shows exponential behavior[35, 36]; and (iii) occurs with higher efficiency and at lower power density ($\approx 10$ W/cm$^2$).

The EL emission peaks at 1.547 eV, which is 93 meV below the monolayer PL in Figure 1d. We assign the shift to different dielectric environments in both experiments, which influence the exciton binding energy due to Coulomb screening. As illustrated in Figure 4b, the spectral position of the PL, recorded directly from the device (1.573 eV), roughly coincidences with that obtained in EL, evidencing that EL arises from an excitonic transition. The large exciton binding energy in TMDCs[37, 38], may thus offer an opportunity for tailoring the emission by engineering of the dielectric environment. Close comparison of EL and PL reveals, that the EL spectrum is slightly broadened towards low energy, which may be attributed to charged excitons[37, 38] being involved in the emission process. However, investigation of more homogeneous samples with narrower linewidths will be necessary to clarify this.

For the future, we envision low-cost, flexible and semi-transparent solar cells that could be deployed on glass facades or other surfaces for energy harvesting. 2D light emitting diodes could lead to new generations of large-area lighting units and transparent, flexible displays. We also expect applications in the emerging field of valleytronics[39].

*Note added:* During the review of this letter we became aware of two similar studies[40, 41].

**Methods**

Device fabrication started by standard electron beam lithographic and metal deposition techniques to produce Ti/Au (7/35 nm) split gate electrodes ($d$ = 460 nm wide gap) on a commercial Si/SiO$_2$ wafer (340 nm oxide thickness). In a second lithography step, 250×250 μm$^2$ large Ti/Au (15/70 nm) bonding pads, connected to the gate electrodes, were patterned. Using plasma-enhanced chemical vapor deposition, a 100-nm-thick Si$_3$N$_4$ layer was uniformly deposited across the sample surface, apart from the bonding pad area, which was covered with a shadow mask. WSe$_2$ was mechanically exfoliated from bulk (Nanosurf Inc.) onto a stack of polymers[34] deposited on top of a sacrificial silicon wafer. The layer thickness of the polymer stack was chosen such that WSe$_2$ monolayers could be identified with an optical microscope. By dissolving the bottom polymer in water, the top polymer with the WSe$_2$ flake was released from the wafer, turned upside down, and placed with micrometer-precision across the gap, such that it overlaps on both sides with the gate electrodes. The top polymer was then dissolved. Finally, the anode and cathode contact electrodes were patterned lithographically, where – for reasons that are explained in the main text – the anode was made of Pd/Au (20/40 nm) and the cathode of Ti/Au (20/40 nm). The nominal widths of the electrostatic n- and p-doping regions between the contact electrodes and the intrinsic (ungated) region of the device are 900 nm each. The sample was annealed for several hours at 380 K in vacuum to remove doping adsorbates and water from the surface.

Photovoltaic response was measured with white light from a halogen lamp (its emission spectrum is presented as SI), that we focused with a microscope objective onto the sample. In advance, we calibrated the optical power density by measuring the transmission through a 10-μm-diameter pinhole. Electroluminescence was collected with a 20× microscope objective and fed via a multimode fiber into a grating spectrometer, equipped with a silicon photodetector array. For better long-term stability, the sample was placed in a high vacuum chamber ($\approx 5\times10^{-6}$ mbar) and kept there during all measurements. The device was found to be functional in ambient air as well.

**Acknowledgements**

We would like to thank Karl Unterrainer for the encouragement, Martin Brandstetter, Michael Krall, and Werner Schrenk for technical assistance, and Emmerich Bertagnolli for providing us access to a Raman spectrometer. The research leading to these results has received funding from the Austrian Science Fund FWF (START Y-539) and the European Union Seventh Framework Programme under grant agreement n°604391 Graphene Flagship.

**Author Contributions**

T.M. conceived the experiment. A.P. fabricated the devices and carried out the measurements. M.F. contributed to the sample fabrication. A.P. and M.F. built the experimental setups. A.P. and T.M. analyzed the data. T.M. prepared the manuscript. All authors discussed the results and commented on the manuscript.

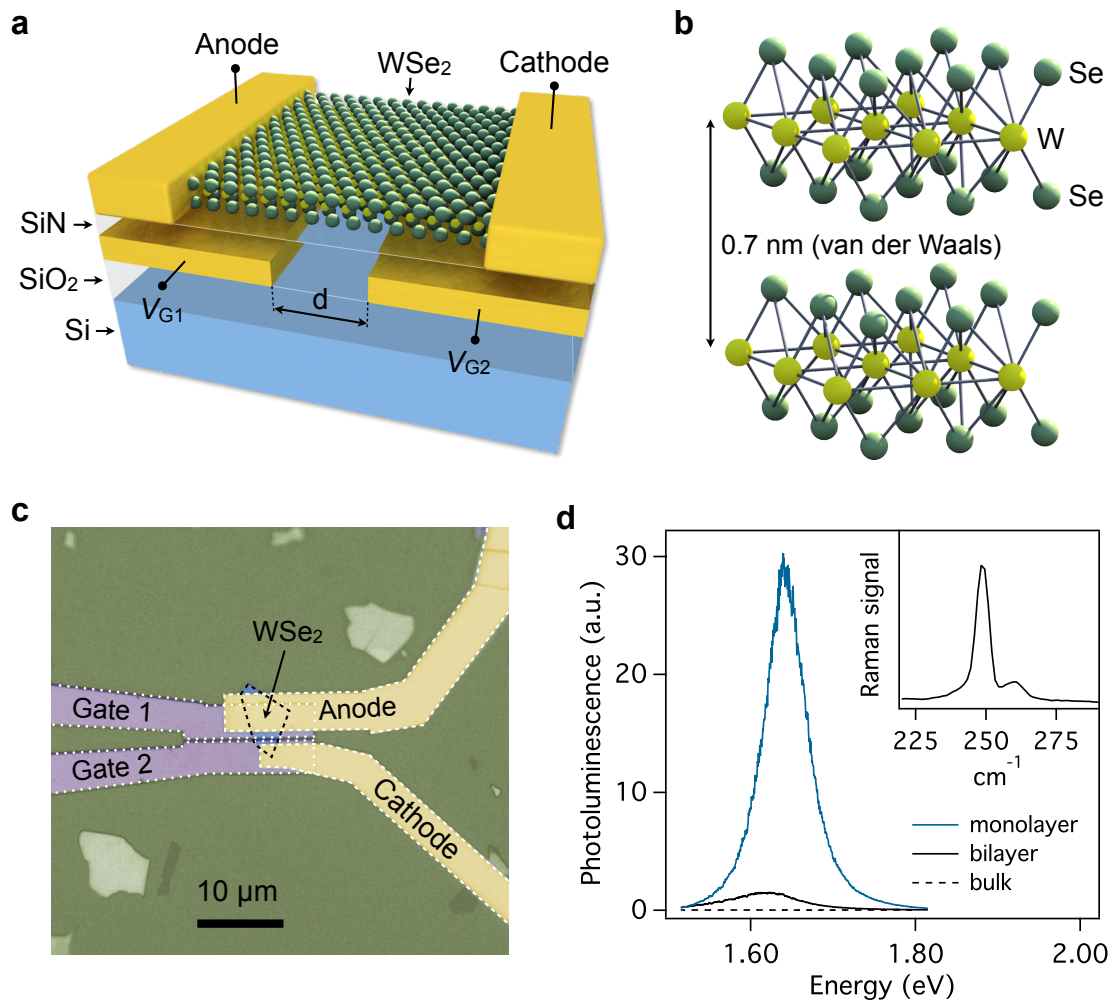

*Figure 1.* **WSe$_2$ monolayer device with split gate electrodes.** (a) Schematic drawing of the device structure. (b) Three-dimensional schematic representation of WSe$_2$. Monolayers are ≈ 0.7 nm thick and are mechanically exfoliated from a bulk crystal. (c) Colored microscope image of the device. In between the gate electrodes and the WSe$_2$ flake there is a 100-nm-thick gate dielectric. The anode electrode is made of Pd/Au and the cathode of Ti/Au. The gap between the gates is 460 nm wide. (d) PL from mono- (solid blue line), bi- (solid black line), and multi-layer (dashed black line) WSe$_2$ flakes. Inset: Raman spectrum of a monolayer.

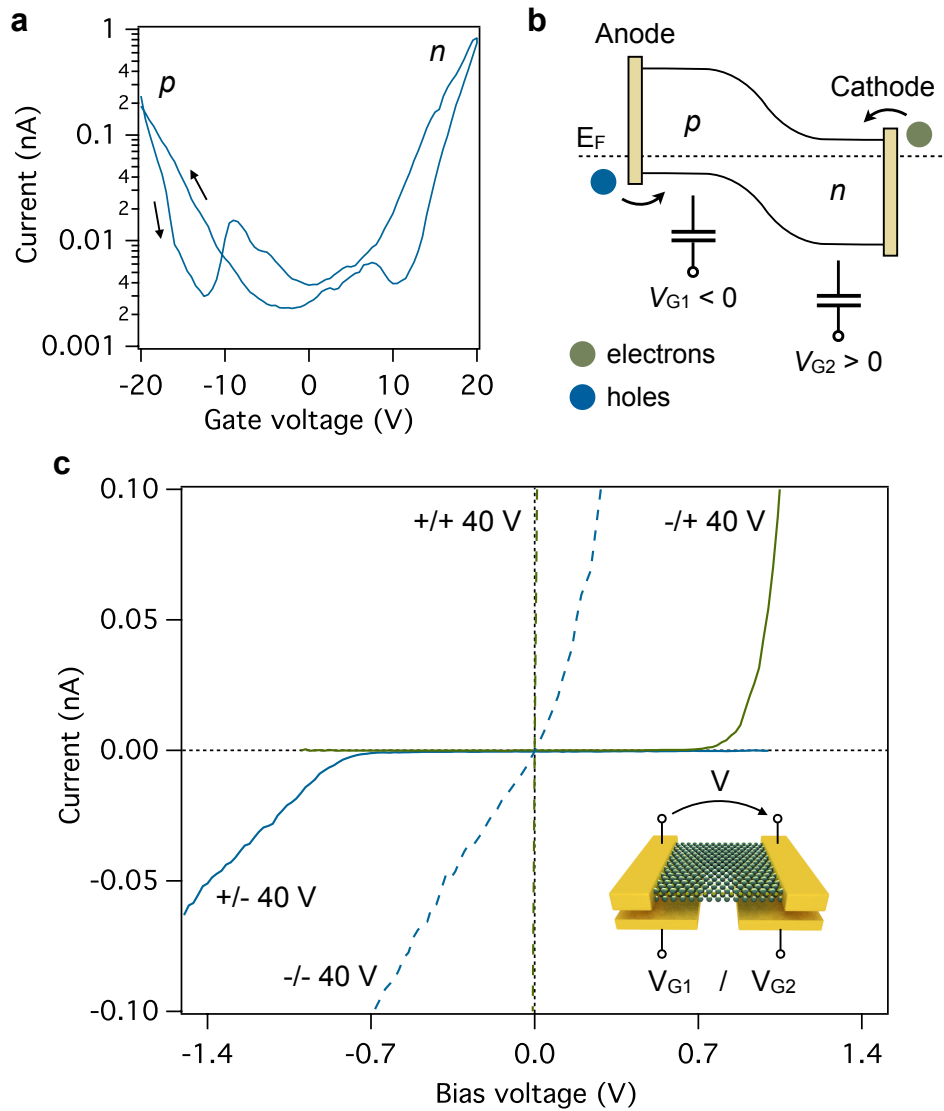

*Figure 2.* **Electrical characterization.** (a) Gate characteristic of the device (0.2 V bias voltage). Both electrons and holes can be injected into the channel. The curve was obtained by scanning the gate voltage from -20 V to 20 V and back. (b) Band diagram when operated as p-n junction diode ($V_{G1} < 0$, $V_{G2} > 0$). Asymmetric contact metallization allows more efficient electron (green) and hole (blue) injection. (c) I-V characteristics of the device in the dark for biasing conditions as shown in the inset: p-n (solid green line), n-p (solid blue line), n-n (dashed green line), p-p (dashed blue line).

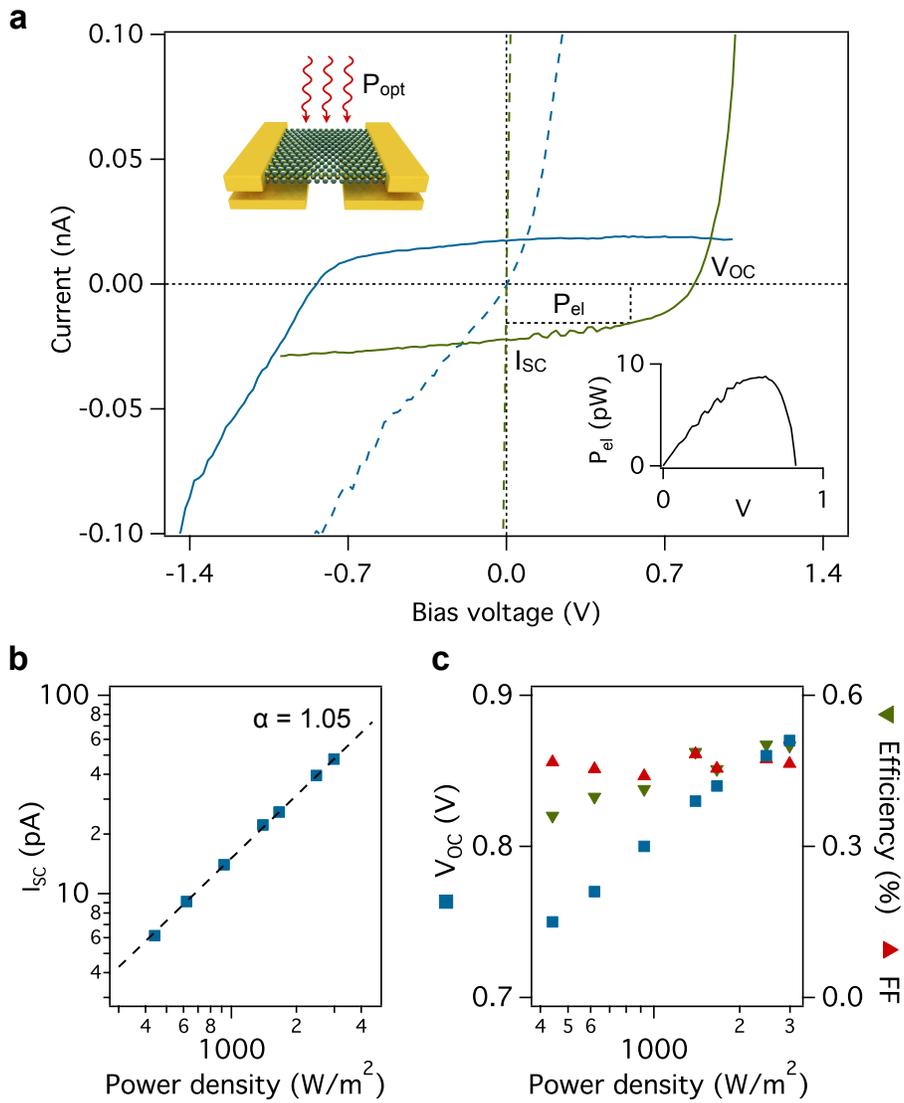

*Figure 3.* **Device operation as solar cell and photodiode.** (a) I-V characteristics of the device under optical illumination with 1400 W/m$^2$. The meaning of the curves is the same as in Figure 2c: p-n (solid green line), n-p (solid blue line), n-n (dashed green line), p-p (dashed blue line). When operated as diode (solid lines), electrical power, $P_{el}$, can be extracted. Inset: $P_{el}$ versus voltage under incident illumination of 1400 W/m$^2$. Maximum power conversion efficiency is obtained for $V_m = 0.64$ V and $I_m = 14$ pA. The dashed rectangle in the main panel shows the corresponding power area. (b) Short-circuit current $I_{SC}$. Symbols: measurement; dashed line: fit of power-law. (c) Open-circuit voltage $V_{OC}$ (green symbols), fill factor $FF$ (red symbols), and power conversion efficiency $\eta_{PV}$ (blue symbols). All parameters are plotted versus incident light intensity.

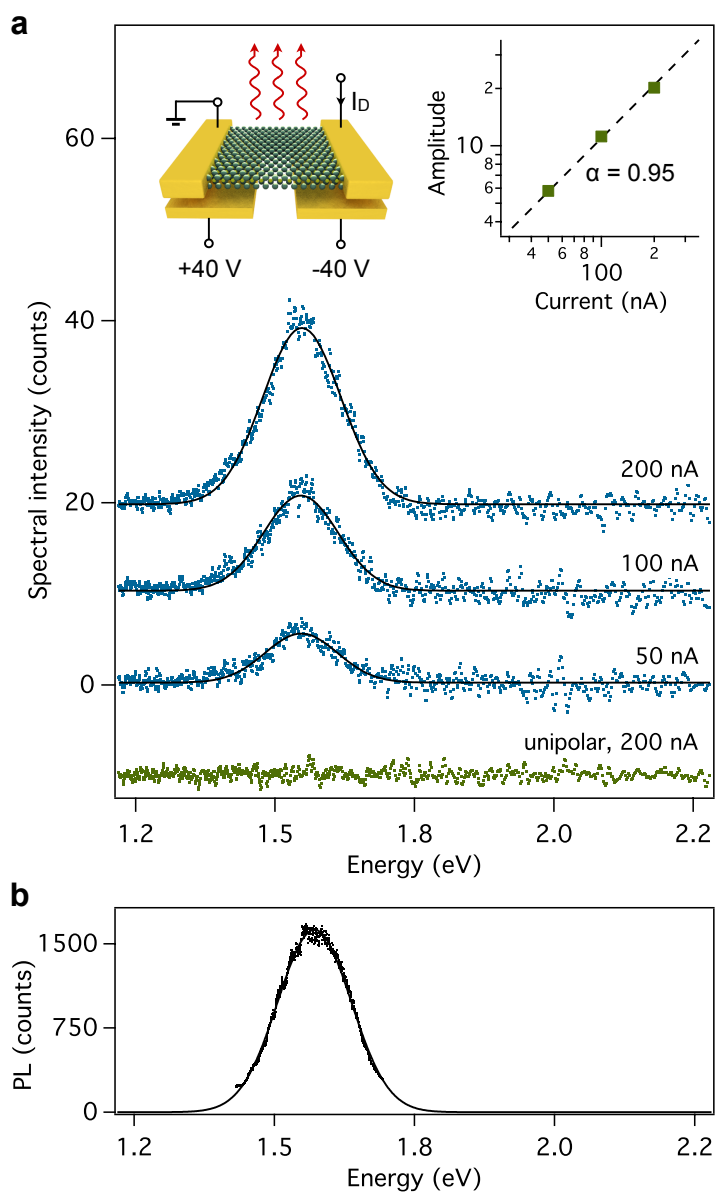

*Figure 4.* **Device operation as light emitting diode.** (a) Electroluminescence emission spectra recorded for gate voltages as shown in the inset and constant currents of 50, 100, and 200 nA, respectively (blue symbols: measurements, black lines: Gaussian fits). Curves are offset for clarity. The green curve demonstrates that no light emission is obtained under unipolar (n-type) conduction. Inset: Emission amplitude versus current on a double-logarithmic scale. Symbols: measurement; dashed line: the data can be fitted by a power-law $I^{\alpha}$ with $\alpha$ close to one ($\alpha$ = 0.95). (b) Photoluminescence recorded from the WSe$_2$ flake on the device. Symbols: measurement; line: Gaussian fit.

*Supplementary Information*

# Solar energy conversion and light emission in an atomic monolayer p-n diode

Andreas Pospischil, Marco M. Furchi, and Thomas Mueller

*Vienna University of Technology, Institute of Photonics, 1040 Vienna, Austria*

### I.) I-V characteristic of WSe$_2$ p-n junction diode in the dark

The Shockley diode equation relates the current $I$ of a p-n junction to the bias voltage $V$,

$$I = I_s\left(e^{V/nV_T} - 1\right),$$

where $I_S$ is the saturation current, $V_T$ is the thermal voltage ($\approx$ 26 mV at $T$ = 300 K), and $n$ is the ideality factor. A more realistic diode model has to take a series resistance $R_S$ – which in our device is associated with the metal/WSe$_2$ contacts and the p- and n-doping regions – into account. An explicit equation for the diode current can then be obtained[†] by using the Lambert $\mathcal{W}$-function

$$I = \frac{nV_T}{R_S}\mathcal{W}\left(\frac{I_S R_S}{nV_T}e^{(V+I_S R_S)/nV_T}\right) - I_S.$$

The model can reproduce our measurement ($V_{G1}$ = -40 V, $V_{G2}$ = 40 V, blue line in Figure S1a) over several orders of magnitude (solid black line), and we extract $I_S$ = 0.02 fA, $n$ = 2.6, and $R_S$ = 95 MΩ. For bias voltages below $\approx$ 0.5 V, the current falls below the noise floor of our measurement equipment (100 fA). For opposite gate biases ($V_{G1}$ = 40 V, $V_{G2}$ = -40 V; Figure S1b), we find a much larger resistance of $R_S$ = 8 GΩ. The diode parameters $I_S$ and $n$, however, are essentially the same ($I_S$ = 0.03 fA, $n$ = 2.6).

### II.) I-V characteristic of WSe$_2$ p-n junction diode under optical illumination

Figure S2a shows I-V characteristics of the p-n junction diode ($V_{G1}$ = -40 V, $V_{G2}$ = 40 V) under optical illumination with power densities varied between 0 and 3000 W/m$^2$. As described in the main text, a slope of the curves at short circuit is observed that can be modeled by a shunt resistance $R_{SH}$. The I-Vs under optical illumination can be obtained from the equivalent circuit in the inset of Figure S2a,

$$I = I_S\left(e^{(V-IR_S)/nV_T} - 1\right) + \frac{V-IR_S}{R_{SH}} - I_L,$$

where $I_L$ is the photocurrent, and $R_S$ = 95 MΩ, $n$ = 2.6, and $I_S$ = 0.02 fA, as deduced above.

---

[†] Banwell, T. & Jayakumar, A. Exact analytical solution for current flow through diode with series resistance. *Electronics Letters* **36**, 291–292 (2000).

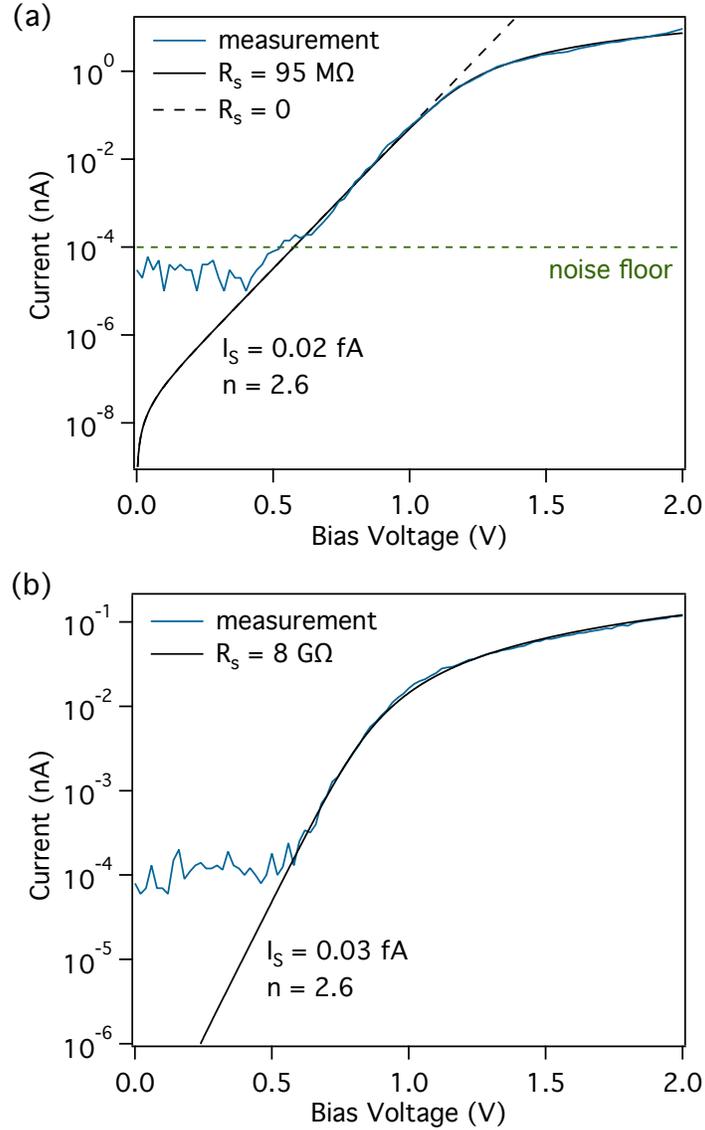

*Figure S1.* I-V characteristic of p-n junction diode under forward bias. Solid blue line: experimental data; solid black line: fit of diode equation with $R_S \neq 0$; dashed black line: fit of diode equation with $R_S$ = 0. $R_S$ denotes the diode serial resistance. The noise floor of the measurement instrument is 100 fA (green dashed line). (a) $V_{G1}$ = -40 V, $V_{G2}$ = 40 V. (b) $V_{G1}$ = 40 V, $V_{G2}$ = -40 V.

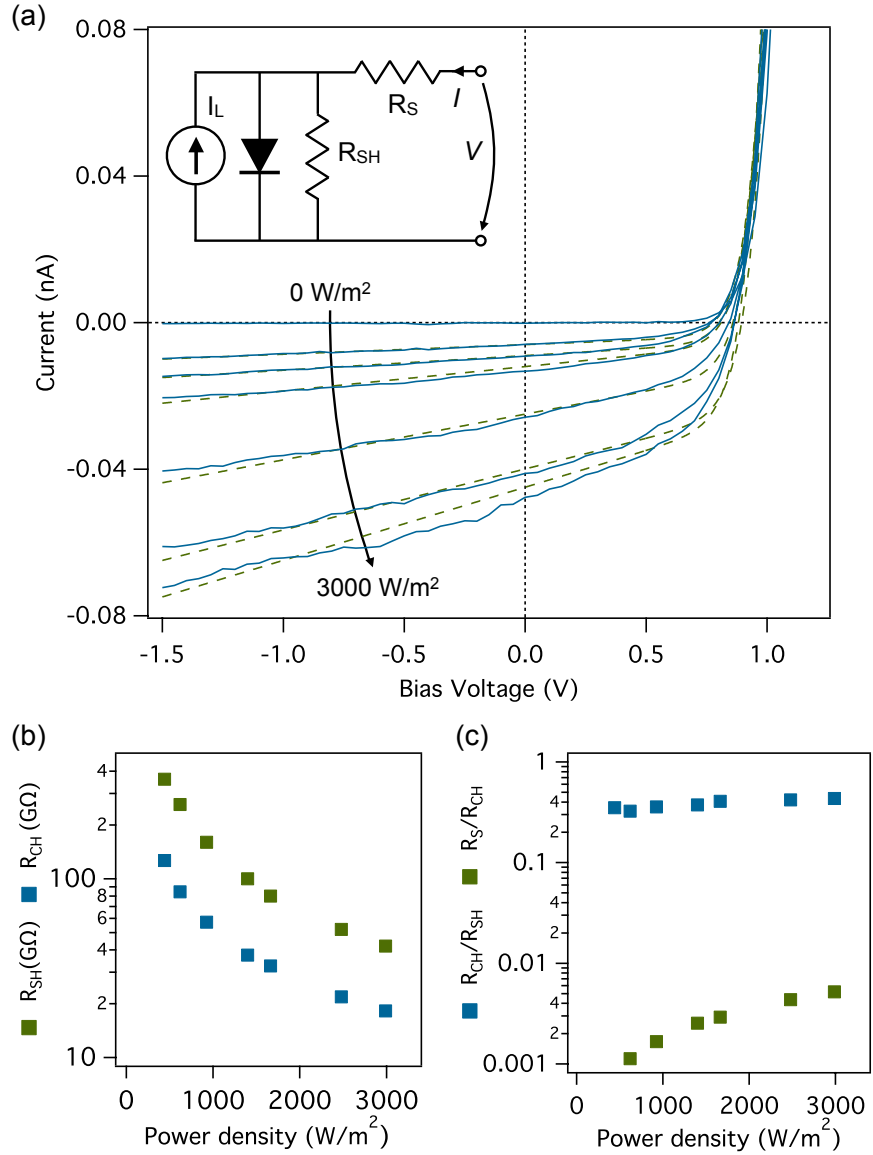

*Figure S2.* (a) I-V characteristic of p-n junction diode ($V_{G1}$ = -40 V, $V_{G2}$ = 40 V) under optical illumination. Solid blue lines: experimental data; dashed green lines: theoretical fits. Inset: Equivalent circuit model of a photovoltaic solar cell. (b) Optical power dependence of characteristic resistance $R_{CH}$ and shunt resistance $R_{SH}$ (extracted from (a)). (c) Normalized series ($R_S/R_{CH}$) and shunt ($R_{CH}/R_{SH}$) resistances, describing dissipation and recombination losses, respectively.

By using the Lambert $\mathcal{W}$-function, $I$ can again be written[‡] as an analytical expression

---

[‡] Jain, A., Sharma, S. & Kapoor, A. Solar cell array parameters using Lambert W-function. *Solar Energy Materials and Solar Cells* **90**, 25–31 (2006).

$$I = \frac{nV_T}{R_S}\mathcal{W}\left(\frac{I_S R_S R_{SH}}{nV_T(R_S+R_{SH})}e^{R_{SH}(V+I_S R_S+I_L R_S)/nV_T(R_S+R_{SH})}\right) + \frac{V-I_S R_{SH}-I_L R_{SH}}{R_S+R_{SH}}.$$

We fit this equation to the measurement data in Figure S2a (dashed lines), extract $R_{SH}$, and plot the results in Figure S2b (green symbols). In the same figure we plot the quantity $R_{CH} = V_m/I_m \approx V_{OC}/I_{SC}$, which is the characteristic resistance of the solar cell (blue symbols). The influence of the shunt on electrical output power can be determined by calculating the power in the absence of $R_{SH}$ minus the power loss in the shunt, $P_{el} \approx V_m I_m - V_m^2/R_{SH}$. A similar analysis for $R_S$ yields $P_{el} \approx V_m I_m - I_m^2 R_S$. In the presence of both series and shunt resistance, we thus obtain for the fill factor of the solar cell

$$FF \approx FF_{ideal}\left(1 - \frac{R_{CH}}{R_{SH}}\right)\left(1 - \frac{R_S}{R_{CH}}\right).$$

From Figure S2c, in which we plot the quantities $R_S/R_{CH}$ and $R_{CH}/R_{SH}$, it becomes apparent that recombination losses (described by $R_{SH}$) severely reduce the fill factor by $\approx$ 40 %, whereas resistive losses ($R_S$) at the contacts are negligible (< 0.5 %).

### III.) Emission spectrum of halogen lamp

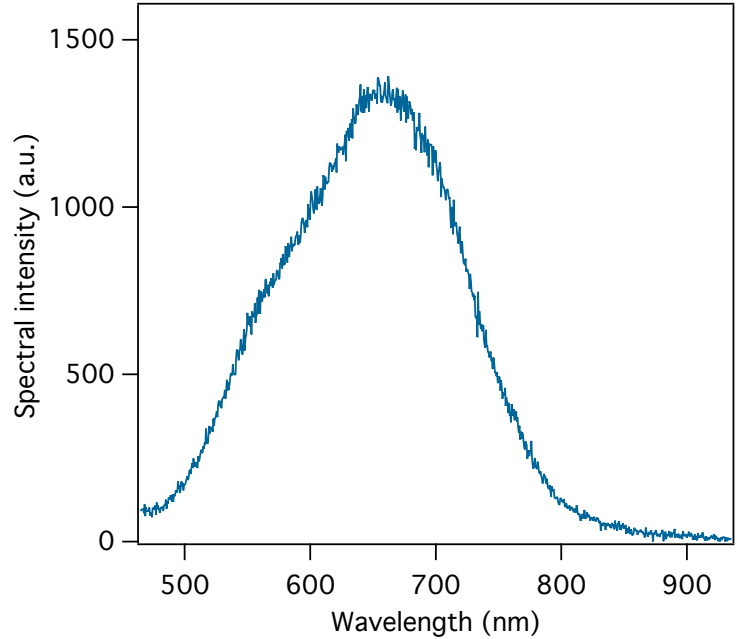

*Figure S3.* Emission spectrum of halogen lamp.

## IV.) Estimation of optical absorption and collection efficiency

The optical absorption in monolayer WSe$_2$ is estimated by integration of the absorption spectrum[§] over the wavelength range of the halogen lamp in Figure S3. Standing wave effects in the device are neglected. The so obtained absorption is ≈ 5 %.

The electroluminescence collection efficiency $\eta_{coll}$ is estimated from the transmission $T$ of all optical components in the beam path (microscope objective, cryostat window, beamsplitter, fiber coupler, etc.) and the integral over the collection angle defined by the numerical aperture *NA* of the objective lens

$$\eta_{coll} = \frac{T}{4\pi}\int_0^{2\pi} d\phi \int_0^{\arcsin(NA)} \sin(\Theta)d\Theta.$$

Modifications of the emission pattern due to the substrate and the electrodes are neglected.

---

[§] Huang, J.-K. *et al.* Large-Area and Highly Crystalline WSe$_2$ Monolayers: from Synthesis to Device Applications. *arXiv:1304.7365* (2013).